\documentclass[12pt]{article}
\begin{document}
\baselineskip=0.5cm  
\begin{center}
\LARGE{A liquid-state theory that remains successful \\ 
in the critical region} \\
\vspace*{0.5cm}
\large{D. Pini, G. Stell} \\ 
\large{\it Department of Chemistry, State University of New York} \\
\large{\it at Stony Brook, Stony Brook, New York 11794--3400, U.S.A.} \\
\vspace*{0.2cm}
\large{N. B. Wilding} \\ 
\large{\it Department of Physics and Astronomy, University of
Edinburgh} \\  
\large{\it Edinburgh EH9 3JZ, U.K.} 
\end{center}
\vspace*{0.3cm}
\begin{abstract}
\baselineskip=0.5cm 
A thermodynamically self-consistent Ornstein-Zernike approximation
(SCOZA)  is applied to a fluid of spherical particles with a pair
potential given by  a hard-core repulsion and a Yukawa attractive tail
$w(r)=-\exp [-z(r-1)]/r$. This potential allows one to take advantage
of the known analytical  properties of the solution to the
Ornstein-Zernike equation for the case  in which the direct correlation
function  outside the repulsive core is given by a linear combination
of two Yukawa  tails and the radial distribution function $g(r)$
satisfies the exact core condition $g(r)=0$ for $r<1$.  The predictions
for the thermodynamics, the critical point, and the coexistence curve
are compared here to other theories and to simulation results. In order
to unambiguously assess the ability of the SCOZA to locate  the
critical point and the phase boundary of the system, a new set  of
simulations has also been performed. The method adopted combines Monte
Carlo and  finite-size scaling techniques and is especially adapted to
deal with critical  fluctuations and phase separation. It is found that
the version of the SCOZA  considered here provides very good  overall
thermodynamics and a remarkably accurate critical point and 
coexistence curve. For the interaction range considered here, given by
$z=1.8$, the critical density and temperature predicted by the theory
agree  with the simulation results to about 0.6\%. 

\end{abstract}
\vfill 
\rule{4cm}{0.5mm} \\
{\small Prepared for the John Barker festschrift 
issue of Molecular Physics.}  \\ 
{\small State University at Stony Brook College 
of Eng. and Appl. Sci. Report No.~754, Jan.~1998.} 
\newpage
\section{Introduction}

After applying their version of thermodynamic perturbation theory to 
square-well and Lennard-Jones fluids, John Barker and Doug Henderson 
characterized it as a ``successful theory of liquids''~\cite{barker}. And so
it was. When tested against simulation results it proved to be impressively
accurate at liquid-state densities and temperatures, unlike some versions
of thermodynamic perturbation theory that had preceded it. And it bypassed
the troubling lack of thermodynamic self-consistency associated with the 
direct use of the radial distribution functions obtained from the 
integral-equation theories then available, as well as yielding thermodynamic
results as good or better than the best results obtainable from such integral
equations. 

These positive features became hallmarks of successful thermodynamic 
perturbation theories for simple fluids and were shared by the 
versions~\cite{mansoori} that followed the Barker and Henderson work as
well  as an alternative perturbative approach set forth somewhat
earlier by Hauge and Hemmer~\cite{hauge} that was based on using the
inverse range of the  attractive interaction rather than its strength
as a perturbation parameter. Integral-equation approaches with improved
self-consistency were also  developed subsequently to yield accurate
liquid-state  thermodynamics~\cite{caccamorev}.

Unfortunately, the accuracy of all these approaches begins to decrease 
substantially as one leaves the liquid-state region located slightly above
the triple point in temperature and follows the liquid-gas coexistence curve
in the density--temperature plane up to the critical region. In particular,
the shape of the coexistence curve and location of the critical point are
not accurately reproduced, nor are related critical parameters. In the case
of the perturbation theories, it is not hard to understand why this is so. 
All of them are mean-field-like in nature, associated with coexistence
curves that are quadratic close to the critical point, whereas the true
coexistence curve is very nearly cubic. That is, in these theories one finds
near the critical point a coexistence curve of the form
\begin{equation}
T_{c}-T \approx A |\rho-\rho_{c}|^{x}, \mbox{\hspace{1cm}} x=2 \, ,
\label{coex}
\end{equation}
where $\rho_{c}$ and $T_{c}$ are the critical values of number density $\rho$
and absolute temperature $T$, and $A$ is a constant. In contrast, in an 
exact treatment, one would expect to find $x$ close to $3$. In these theories
the resulting $T_{c}$ is usually more than 5\% too high and the critical
compressibility factor $(P/\rho k_{B}T)_{c}$ is usually more than 10\% too
high. Here $P$ is the pressure, and $k_{B}$ is the Boltzmann constant. 

The thermodynamics associated with the radial distribution function $g(r)$
obtained form various integral-equation approaches cannot be so neatly
categorized. However, in the cases in which there are substantial 
discrepancies between the several paths available for obtaining thermodynamics
from $g(r)$, the most reliable and accurate coexistence behavior
is often obtained from evaluating the thermodynamics through the excess 
internal energy expressed in terms of an integral over the pair potential
$w(r)$ weighted by $g(r)$. For continuum-fluid models the resulting critical
behavior is typically mean-field like in the cases that we have 
studied, and thus subject to the same deficiencies as one approaches 
the critical region. In some integral-equation approaches that have been
developed in order to insure a certain degree of thermodynamic consistency,
the description of the critical region and of the phase diagram appears
to be more problematic: 
for instance, the modified hypernetted chain (MHNC) theory~\cite{rosenfeld}
is indeed able to predict quite satisfactorily the
liquid and the vapor branches of the coexistence curve of a simple fluid 
at low enough temperature, but it fails to converge close to the critical
point, so that the two branches remain unconnected, and the position of the 
critical point is not given directly by the theory, but must be determined
by extrapolation~\cite{caccamoyuk,caccamoljrhnc}. The same kind of 
behavior~\cite{caccamoyuk,caccamoljhmsa}
is found also for the HMSA integral equation [the acronym coming from the 
fact that the theory~\cite{zerah} interpolates between the hypernetted chain 
(HNC) and the soft mean spherical approximation (SMSA)]. 

The self-consistent Ornstein-Zernike approximation (SCOZA) we consider here 
is not mean-field-like, and it remains highly accurate as one goes from
liquid-state conditions to critical-point conditions. In particular the power
$x$ in Eq.~(\ref{coex}) was recently shown analytically to be given
in the SCOZA by exactly $20/7$~\cite{scozacrit}. And as we discuss in this
paper, in the hard-core Yukawa fluid (HCYF) $T_{c}$ appears to be within 
0.6\% of its value as estimated by our simulation results. (Similarly, in
recent three-dimensional lattice-gas studies~\cite{dickman,scoza1} the SCOZA
$T_{c}$ was found to be within 0.2\% of its estimated exact value). 
As described elsewhere~\cite{scozacrit,scoza1} the scaling behavior of the
SCOZA thermodynamics is somewhat different from the simple scaling one expects
to see in the exact thermodynamics, although those differences only begin
to appear clearly when $\rho$ and $T$ are within less than 1\% of their 
critical values. Closer to the critical point, the effective exponents defined
above $T_{c}$ approach spherical-model values as the critical point is 
approached, whereas the exponents defined below $T_{c}$ do not. 
The exponents are discussed in Sec.~3.

The SCOZA was proposed some time ago by H\o ye and Stell~\cite{hoye1,twoyuk3}
but fast and accurate algorithms for evaluating its thermodynamic predictions
were developed only recently~\cite{dickman,scoza1,scoza2}. A sharp 
assessment of its accuracy for the HCYF could not be made on the basis 
of existing simulations, and for that reason our study here includes new
Monte Carlo (MC) results exploiting finite-size scaling (FSS)
techniques~\cite{wilding}. 

We have chosen the HCYF pair potential as the first of the continuum-fluid
potentials to be considered in our studies of the SCOZA for several reasons.
First, it embodies the two key features one requires in an off-lattice pair
potential in order to consider both the liquid state and liquid-gas 
criticality---a highly repulsive core and an attractive well. Second, the 
HCYF proves to be particularly convenient to analyze using the SCOZA
(the square-well fluid is far less convenient in this regard). Third,
the functional form of the hard-core Yukawa potential makes it appropriate
as a generic solvent-averaged interaction potential between polyelectrolytes
and colloids as well as a generic simple-fluid pair potential. For 
this reason it seems particularly useful to have an accurate theory for 
both the structure and thermodynamics of the HCYF, which has already been the
subject of a number of previous studies. We shall make contact with several 
of those here.

The paper is organized as follows: in Sec.~2 we describe the theory and
present some details of the method for the system under study, in Sec.~3 our
results are shown and a comparison with other theories and simulation
results is made, and in Sec.~4 our conclusions are drawn. The treatment
of the hard-sphere gas and the main features of the MC-FSS simulation method
are summarized respectively in Appendix~A and Appendix~B.

\section{Theory}

Here we consider a fluid of spherical particles interacting via a two-body 
potential $v(r)$ which is the sum of a singular repulsive hard-sphere 
contribution and an attractive tail $w(r)<0$. The expression for $v(r)$ is then
\begin{equation}
v(r)=\left \{ \begin{array}{ll}
 +\infty \mbox{\hspace{1cm}}   & r < 1 \\
                               &       \\
 w(r) & r>1 \, ,
\end{array}
\right.
\label{pot}
\end{equation}
where the hard-sphere diameter has been set equal equal to unity. 
As is customary in 
integral equation theories of fluids, the present approach introduces
an approximate closure relation for the direct correlation function $c(r)$
which, once supplemented with the exact Ornstein-Zernike equation involving 
$c(r)$ and the radial distribution function $g(r)$, yields a closed theory 
for the thermodynamics and the correlations of the system under study. 
The basic requirement we want to incorporate in the SCOZA is the consistency 
between the compressibility and internal energy route to the thermodynamics. 
According to the compressibility route, the thermodynamics stems from the
reduced compressibility $\chi_{\rm red}$ as determined by the sum rule
\begin{equation}
\chi_{\rm red}=\frac{1}{1-\rho \widehat{c}(k=0)} \, ,
\end{equation}
where $\widehat{c}(k)$ is the Fourier transform of the direct correlation 
function and $\rho $ is the number density of the system. In the internal 
energy route the key to the thermodynamics is instead provided by the excess
internal energy as given by the integral of the interaction weighted 
by the radial distribution function:
\begin{equation}
u=2\pi \rho^2 \int_{1}^{+\infty}\! dr \, r^2 w(r)g(r) \, ,
\label{energy}
\end{equation}
where $u$ is the excess internal energy per unit volume and we have taken 
into account that $g(r)$ vanishes for $r<1$ due to the hard-core repulsion.
In the following we will refer to the ``excess internal energy'' simply as the
``internal energy''. 
If $\chi_{\rm red}$ and $u$ come from a unique Helmholtz free energy it
is straightforward to find that one must have
\begin{equation}
\frac{\partial}{\partial \beta}\left(\frac{1}{\chi_{\rm red}}\right)=
\rho \frac{\partial^2 u}{\partial \rho^2} \, ,
\label{consist}
\end{equation}
where $\beta=1/(k_B T)$, $T$ being the absolute temperature, and $k_B$ 
the Boltzmann constant. While this relation is of course satisfied by the
exact compressibility and internal energy, this is not the case 
with those predicted by most integral equation theories. In order 
to cope with this lack of thermodynamic consistency, we consider 
the following closure to the Ornstein-Zernike equation:
\begin{equation}
\left\{
\begin{array}{ll}
g(r)=0                     & r<1 \, , \\
                           & \\
c(r)=c_{\rm HS}(r)+K(\rho, \beta)w(r) \mbox{\hspace{0.3cm}} & r>1  \, ,
\end{array}
\right.
\label{closure}
\end{equation}
where $c_{\rm HS}(r)$ is the direct correlation function of the hard-sphere
fluid, and $K(\rho, \beta)$ is a function of the thermodynamic state of the
system. In Eq.~(\ref{closure}) the approximation clearly lies in the simple
form of $c(r)$ outside the repulsive core. 
The closure above resembles the one used in the approximation known as both
the lowest-order gamma-ordered approximation (LOGA)~\cite{stell} and
the optimized random phase approximation (ORPA)~\cite{andersen}.
However, while in the LOGA/ORPA one has $K(\rho, \beta)\equiv -\beta$,
in Eq.~(\ref{closure}) $K(\rho, \beta)$ is not fixed {\it a priori}, 
but instead must be
determined so that the thermodynamic consistency condition~(\ref{consist})
is satisfied. This gives rise to a partial differential equation (PDE) for
the function $K(\rho ,\beta)$, provided an expression for the hard-sphere 
direct correlation function $c_{\rm HS}(r)$ is given.
The most popular parameterization for $c_{\rm HS}(r)$ in the fluid region
is due to Verlet and Weis~\cite{verlet}. Another choice that yields
the same thermodynamics as Verlet-Weis, and that we find convenient in view 
of the calculations performed in this work, is originally due to 
Waisman~\cite{waisman}. It was subsequently extended analytically by H\o ye
and Stell~\cite{oneyuk} and explored in some detail by Henderson and
coworkers~\cite{blum}. It amounts to assuming
that the function $c_{\rm HS}(r)$ outside the repulsive core has 
a one-Yukawa form, so that for the hard-sphere system we have:
\begin{equation}
\left\{
\begin{array}{ll}
g_{\rm HS}(r)=0 & r<1 \, , \\
                &     \\
c_{\rm HS}(r)=K_1 \displaystyle{\frac{\exp[-z_1 (r-1)]}{r}} 
\mbox{\hspace{0.3cm}} & r>1 \, .
\end{array}
\right.
\label{waisman}
\end{equation}
The Ornstein-Zernike equation supplemented by Eq.~(\ref{waisman}) can be 
solved analytically in terms of the amplitude $K_1$ and the inverse range
$z_1$ of $c_{\rm HS}(r)$. These can be in turn determined as a function
of the density by requiring, as in the Verlet-Weis parameterization, 
that both the compressibility and the virial route to the thermodynamics
give the Carnahan-Starling equation of state. The basic features of the
calculation are recalled in Appendix~A. 

A considerable, although purely technical, simplification in the closure
scheme outlined above based on Eqs.~(\ref{consist}),~(\ref{closure}) occurs 
when also the attractive potential $w(r)$ in Eq.~(\ref{pot}) is given 
by a Yukawa function, i.e. when one has 
\begin{equation}
w(r)=-\frac{\exp[-z (r-1)]}{r} \, ,
\label{yuk}
\end{equation}
$z$ being the inverse range of the potential. From 
Eq.~(\ref{waisman}) it is then immediately seen that Eq.~(\ref{closure})
becomes
\begin{equation}
\left\{
\begin{array}{ll}
g(r)=0   &  r<1 \, , \\
         &       \\
c(r)=K_1 \displaystyle{\frac{\exp[-z_1 (r-1)]}{r}}+K_2 
\displaystyle{\frac{\exp[-z_2 (r-1)]}{r}} \mbox{\hspace{0.3cm}} & r>1 \, ,
\end{array}
\right.
\label{twoyuk}
\end{equation}
where $K_2$ and $z_2$ are the quantities referred to as $K$ and $z$ 
in Eq.~(\ref{closure}),~(\ref{yuk}), and
$K_1$, $z_1$ are known function of the density. It is now possible to 
take advantage of the fact that for the Ornstein-Zernike equation supplemented
by the closure~(\ref{twoyuk}) extensive analytical results have been 
determined~\cite{twoyuk1,twoyuk2,twoyuk3}. If both $K_1$ and $K_2$
are given, as in the LOGA/ORPA, this enables one to solve Eq.~(\ref{twoyuk})
altogether~\cite{konior1,konior2,konior3}. More generally, irrespective
of the form of $K_1$ and $K_2$, a prescription can be found to determine
the reduced compressibility $\chi_{\rm red}$ as a function of the density
$\rho$ and the internal energy per unit volume $u$, which can be used 
in Eq.~(\ref{consist}) to
obtain a closed PDE. A similar procedure for the same potential considered
here was adopted in a previous work~\cite{scoza2}, where however the 
hard-sphere contribution to the direct correlation function $c_{\rm HS}(r)$
outside the core was not taken into account, so that $c(r)$ was given by 
a simple one-Yukawa tail. This further simplifies the theory, but implies
that the description of the hard-sphere fluid coincides with that of the
PY approximation, which as is well known is not very satisfactory 
at high density. This defect becomes more and more severe as the range of
the attractive interaction decreases, and can considerably affect the 
phase diagram predicted by the theory,
unless some more-or-less {\it ad hoc}
procedure is adopted to correct the hard-sphere thermodynamics. In order to
incorporate a better treatment of the hard-sphere fluid into the 
theory one can turn to the two-Yukawa form for $c(r)$ of 
Eq.~(\ref{twoyuk}), whose use in the consistency condition~(\ref{consist})
we are now going to illustrate in some detail. In the following we will
exploit the results determined in Refs.\cite{twoyuk1,twoyuk2,twoyuk3},
which will be respectively referred to as~I,~II,~III. Let us introduce 
the packing fraction $\xi=\pi\rho/6$ and the quantity
\begin{equation}
f=(1-\xi)\sqrt{\frac{1}{\chi_{\rm red}}} \, ,
\label{f}
\end{equation}
which is the square root of the quantity referred to as $A$ in I,~II,~III. 
Eq.~(\ref{consist}) becomes 
\begin{equation}
\frac{2f}{(1-\xi)^2}\left(\frac{\partial f}{\partial u}\right)_{\! \rho} \!
\left(\frac{\partial u}{\partial \beta}\right)_{\! \rho} = 
\rho \left(\frac{\partial^2 u}{\partial \rho^2}\right)_{\! \beta} \, .
\label{consist2}
\end{equation}
To obtain a PDE for $u$ we need to express $f$ as a function of $\rho$
and $u$ in Eq.~(\ref{consist2}). From Eq.~(II.14) it is found that $f$
can be written as
\begin{equation}
f=-\frac{(z_{1}^{2}-z_{2}^{2})+4\sqrt{q}\, (\gamma_2-\gamma_1)}
{4\left[ (z_1/\!z_2)\, \gamma_{2}-(z_2/\!z_1)\, \gamma_{1}\right]} - 
\frac{z_{1}^{2}-z_{2}^{2}}{z_{1}z_{2}} \, 
\frac{\gamma_{1}\gamma_{2}(\gamma_{2}-\gamma_{1})}
{\left[ (z_1/\!z_2) \, \gamma_{2}-(z_2/\!z_1) \, \gamma_{1}\right]^2} \, ,
\label{f2}
\end{equation}
where we have set
\begin{equation}
q=\frac{(1+2\xi)^2}{(1-\xi)^2 } \, .
\label{q}
\end{equation}
The quantities $\gamma_1$ and $\gamma_2$ are given by Eq.~(II.5)
\begin{eqnarray}
\gamma_1 & = & 2-\sqrt{q}-\frac{U_1}{U_0} \label{gamma1} \, , \\
\gamma_2 & = & 2-\sqrt{q}-\frac{W_1}{W_0} \label{gamma2} \, .
\end{eqnarray}
The ratios $U_1/U_0$ and $W_1/W_0$ depend on the integrals
\begin{equation}
I_{i}=4\pi \rho \int_{1}^{+\infty}\! dr\,  r\exp[-z_{i}(r-1)]\, g(r)  
\mbox{\hspace{1cm}} (i=1,2).
\label{int}
\end{equation}
>From Eq.~(I.35) it is found in fact
\begin{equation}
\frac{W_1}{W_0} = \frac{4+2z_{2}-z_{2}^{2}}{2(2+z_{2})} \, 
\frac{\tau_{2}I_{2}-1}{\sigma_{2}I_{2}-1} \, ,
\label{w}
\end{equation}
and the corresponding relation with $W_1/W_0$ replaced by $U_1/U_0$ and the
index $2$ changed to $1$. The quantities $\tau_{i}$ and $\sigma_{i}$ 
depend only on $z_{i}$ and are given by Eq.~(I.34):
\begin{eqnarray}
\sigma_{i} & = & \frac{1}{2z_{i}}\left[\frac{z_{i}-2}{z_{i}+2}+\exp(-z_{i})
\right] \, , \label{sigma} \\
\tau_{i} & = & \frac{1}{2z_{i}}\left[\frac{z_{i}^{2}+2z_{i}-4}
{4+2z_{i}-z_{i}^2}+\exp(-z_{i})\right]\, ,
\label{tau}
\end{eqnarray}
with $i=1,2$. From the expression of the potential~(\ref{yuk}) it is 
readily seen
that $I_2$ is directly related to the internal energy per unit volume 
$u$ given by Eq.~(\ref{energy}):
\begin{equation}
u=-\frac{1}{2}\, \rho I_2 \, .
\label{u}
\end{equation}
Eqs.~(\ref{gamma2}),~(\ref{w}),~(\ref{u})
allow then to express $\gamma_{2}$ explicitly
as a function of $\rho$ and $u$:
\begin{equation}
\gamma_{2}=2-\sqrt{q}-\frac{4+2z_{2}-z_{2}^{2}}{2(2+z_{2})} \,
\frac{2\tau_{2}\, u+\rho}{2\sigma_{2}\, u+\rho} \, .
\label{gamma22}
\end{equation}
We now need $\gamma_{1}$ as a function of $\rho$ and $u$. This is less 
straightforward than for $\gamma_{2}$, since the integral $I_1$ does not 
have any direct
thermodynamic meaning, the exponential in $I_1$ being related to the
tail of the direct correlation function of the hard-sphere gas. We have then
to make use of some further results determined in I--III. 
>From Eq.~(I.36) it is found that the amplitudes $K_1$, $K_2$ of the Yukawa
functions in the closure~(\ref{twoyuk}) can be expressed in terms of the
above introduced quantities $U_0$, $U_1$, $W_0$, $W_1$. One has
\begin{equation}
K_{1} = \frac{2(z_{1}+2)^{2}\sigma_{1}^{2}}{3\xi z_{1}^{2}}\, U_{0}
\left[\, \frac{U_1}{U_0}-\alpha_{1}\right]^{2} \, ,
\label{k1}
\end{equation}
where $\alpha_1$ is given by Eq.~(I.37):
\begin{equation}
\alpha_{1} = \frac{(4+2z_{1}-z_{1}^{2})\tau_{1}}{2(2+z_{1})\sigma_{1}} \, ,
\label{alpha}
\end{equation}
and the corresponding equations with the index $1$ replaced by $2$ and $U_0$, 
$U_1$ replaced by $W_0$, $W_1$. Let us now introduce the quantities $x$, $y$
given by
\begin{eqnarray}
x & = & \sqrt{q}-\frac{z_{1}^{2}}{4\gamma_{1}} \, , \label{x} \\
y & = & \sqrt{q}-\frac{z_{2}^{2}}{4\gamma_{2}} \, . \label{y}
\end{eqnarray}
>From Eq.~(III.30) one has
\begin{eqnarray}
U_{0} & = & \frac{4}{z_{1}^{2}}\, p \, (\sqrt{q}-x)^{2} \, , \label{u0} \\
W_{0} & = & \frac{4}{z_{2}^{2}}\, s \, (\sqrt{q}-y)^{2} \, . \label{w0} 
\end{eqnarray}
where $p$ and $s$ must satisfy Eq.~(II.39) (in the notation of II one has 
$x\equiv u_{q1}/u_{q0}$, $y\equiv w_{q1}/w_{q0}$, $p\equiv u_{q0}$, 
$s\equiv w_{q0}$):
\begin{equation}
\left\{
\begin{array}{lll}
p+s+\displaystyle{\frac{4s}{z_{1}^{2}-z_{2}^{2}}}
(y-x)^{2} & = & \frac{1}{4}z_{1}^{2}-x^2 
\, , \\
p+s-\displaystyle{\frac{4p}{z_{1}^{2}-z_{2}^{2}}}
(y-x)^{2} & = & \frac{1}{4}z_{2}^{2}-y^2
\, .
\end{array}
\right.
\label{syst}
\end{equation}
Eq.~(\ref{syst}) is readily solved for $p$ and $s$ to give 
\begin{equation}
p=-\frac{z_{1}^{2}-z_{2}^{2}}{64(y-x)^{4}}\left\{4z_{2}^{2}(y-x)^{2}-
16y^{2}(y-x)^{2}-(z_{1}^{2}-z_{2}^{2})
\left[z_{1}^{2}-z_{2}^{2}+4(y^{2}-x^{2})\right]\right\} \, ,
\label{p}
\end{equation}
and the expression for $s$ is obtained by exchanging $z_1$, $z_2$ and 
$x$, $y$ in the r.h.s. of Eq.~(\ref{p}). If Eqs.~(\ref{gamma1}), (\ref{x}),
(\ref{u0}), (\ref{p}) are used in Eq.~(\ref{k1}) we finally obtain
\begin{eqnarray}
\lefteqn{ \left[ 4(2-\sqrt{q}-\alpha_{1})(\sqrt{q}-x)-z_{1}^{2}\right]^{2}
\left\{4z_{2}^{2}(y-x)^{2}-16y^{2}(y-x)^{2}\right.} \nonumber \\
& & \mbox{} \left. -(z_{1}^{2}-z_{2}^{2})
\left[z_{1}^{2}-z_{2}^{2}+4(y^{2}-x^{2})\right] \right\} 
= - \frac{384\, \xi z_{1}^{4}}{(z_{1}+2)^{2}(z_{1}^{2}-z_{2}^{2})
\sigma_{1}^{2}}\, K_{1}(y-x)^{4} \, ,
\label{msa}
\end{eqnarray}
and a similar equation obtained by exchanging the indices $1$ and $2$ and
the quantities $x$, $y$. We recall that in Eq.~(\ref{msa}) $K_1$, $z_1$, 
$\sigma_1$, and $\alpha_1$ are known functions of the density $\rho$ 
which refer to the hard-sphere system. For given values of $\rho$ and $u$,
Eqs.~(\ref{gamma22}),~(\ref{y}) allow one to determine $y$. Eq.~(\ref{msa})
can then be solved numerically with respect to $x$ to obtain $\gamma_1$ 
via Eq.~(\ref{x}).
This solves the problem of determining $\gamma_1$ in terms of $\rho$ and $u$.
The partial derivative $(\partial f/\partial u)_{\rho}$ that appears in 
Eq.~(\ref{consist2}) can then be determined as
\begin{equation}
\left(\frac{\partial f}{\partial u}\right)_{\!\rho} = 
\left(\frac{\partial f}{\partial \gamma_{1}}\right)_{\!\rho}
\left(\frac{\partial \gamma_{1}}{\partial u}\right)_{\!\rho} +
\left(\frac{\partial f}{\partial \gamma_{2}}\right)_{\!\rho}
\left(\frac{\partial \gamma_{2}}{\partial u}\right)_{\!\rho} \, ,
\label{dfdu}
\end{equation}
where $(\partial \gamma_{2}/\partial u)_{\rho}$ is calculated explicitly 
by Eq.~(\ref{gamma22}), while $(\partial \gamma_{1}/\partial u)_{\rho}$ 
must be determined as the derivative of the function implicitly defined by 
Eq.~(\ref{msa}). If we write Eq.~(\ref{msa}) as $F(x,y,\rho)=0$, it is found
straightforwardly that Eq.~(\ref{consist2}) takes the form
\begin{equation}
B(\rho, u) \frac{\partial u}{\partial \beta} = 
C(\rho, u) \frac{\partial^{2} u}{\partial \rho^{2}} \, ,
\label{pde}
\end{equation}
where the functions $B(\rho, u)$ and $C(\rho, u)$ are given by the following
expressions:
\begin{eqnarray}
B(\rho, u) & = & \frac{2f}{(1-\xi)^{2}}\, 
\frac{\partial \gamma_{2}}{\partial u}
\left[\frac{\partial f}{\partial \gamma_{2}} \frac{\partial F}{\partial x}
\frac{\partial x}{\partial \gamma_{1}} -
\frac{\partial f}{\partial \gamma_{1}} \frac{\partial F}{\partial y}
\frac{\partial y}{\partial \gamma_{2}}\right] \, ,  \label{b} \\
C(\rho, u) & = & \rho \frac{\partial F}{\partial x} \,
\frac{\partial x}{\partial \gamma_{1}} \, . \label{c} 
\end{eqnarray}
All the partial derivatives in Eqs.~(\ref{b}),~(\ref{c}) are
calculated at constant $\rho$ and can be determined by Eqs.~(\ref{f2}), 
(\ref{gamma22}), (\ref{x}), (\ref{y}), (\ref{msa}). The resulting expressions
are then evaluated as a function of $\rho$ and $u$ via the procedure 
described above. The same procedure also allows one to determine the reduced
compressibility as $1/\chi_{\rm red}=f^{2}/(1-\xi)^{2}$ once $f$ has been 
obtained from Eq.~(\ref{f2}). The PDE~(\ref{pde}) is a non-linear diffusion
equation that must be integrated numerically. To prevent the occurrence 
of any numerical instability, especially in the critical and sub-critical 
region, we have adopted an implicit finite-differences algorithm~\cite{ames} 
tailored to equations that, although globally non-linear, depend on 
the partial derivatives of the unknown function in a linear fashion like 
Eq.~(\ref{pde}). 
The integration with respect to $\beta$ starts at $\beta=0$ and goes down 
to lower and lower temperatures. Before each integration step Eq.~(\ref{msa})
is solved numerically and the coefficients $B(\rho, u)$, $C(\rho, u)$ 
are determined. The density $\rho$ ranges in a finite interval 
$(0, \rho_{0})$, whose high-density boundary has been typically set at 
$\rho_{0}=1$. The initial condition can be determined by taking into account
that at $\beta=0$ the radial distribution 
function coincides with that of the hard-sphere gas. From Eqs.~(\ref{energy})
and~(\ref{yuk}) one has then 
\begin{equation}
u(\rho, \beta=0) = -2\pi\rho^{2} \int_{0}^{+\infty}\! dr\, r \exp[-z_{2}(r-1)]
\, g_{\rm HS}(r) \mbox{\hspace{1cm} for every $\rho$} \, ,
\label{init}
\end{equation}
where $g_{\rm HS}(r)$ is obtained in the present scheme by the 
closure~(\ref{waisman}). For such a closure, as shown in Appendix~A, both 
$U_0$ and $U_1$ in Eq.~(\ref{gamma1}) can be determined analytically as
a function of $\rho$, thus providing $\gamma_{1}(\rho)$ at $\beta=0$. 
This allows one to obtain $u$ in Eq.~(\ref{init}) analytically as well: 
in fact, one can solve Eq.~(\ref{f2}) for $\gamma_{2}$ as a function of 
$\gamma_{1}$, $f$, and $\rho$, where $f$ is readily obtained by using 
the Carnahan-Starling expression of $\chi_{\rm red}$ in Eq.~(\ref{f}). 
Once $\gamma_{2}$ is known, Eq.~(\ref{gamma22}) is solved with respect to $u$.
It must be noted that 
solving Eq.~(\ref{f2}) for $\gamma_{2}$ gives two branches, so attention
must be paid in order to single out the branch that actually corresponds to
the physical solution. 
We also need two boundary conditions at $\rho=0$ and $\rho=\rho_{0}$. 
>From Eq.~(\ref{energy}) one has immediately
\begin{equation}
u(\rho =0, \beta)=0 \mbox{\hspace{1cm} for every $\beta$} \, .
\label{boundlow}
\end{equation}
At high density we instead make use of the so-called high-temperature 
approximation (HTA), according to which the excess Helmholtz free energy
per unit volume is determined via Eq.~(\ref{init}) for every temperature. 
In the fluid region of the phase diagram this of course is not exact unless
$\beta=0$, but it becomes more and more accurate as the density of the system
is increased~\cite{packing}, so we expect that for a given sweep along the
$\beta$-axis the results will not differ appreciably from what would be
obtained using an hypothetical exact boundary condition, provided the boundary 
$\rho_{0}$ is located at sufficiently high density. We used the HTA at
$\rho =\rho_{0}$ for the reduced compressibility. This yields via
Eq.~(\ref{consist}) the boundary condition
%
\begin{equation}
\frac{\partial^{2} u}{\partial \rho^{2}}(\rho_{0},\beta) =
\frac{\partial^{2} u}{\partial \rho^{2}}(\rho_{0},\beta\!=\!0)
\mbox{\hspace{1cm} for every $\beta$} \, .
\label{boundhigh1}
\end{equation}
%
%
%
We have checked that the output of the numerical integration of 
Eq.~(\ref{pde}) is quite insensitive to the specific choice 
of the high-density boundary condition.
Moreover, for $\rho_{0}\simeq 1$ moving the boundary condition 
to higher densities also leaves the results unaffected. 
Eq.~(\ref{f}) shows that to be physically meaningful, the quantity $f$ 
has to be non-negative. On the other hand, below the critical temperature 
the solution of Eq.~(\ref{pde}) does not satisfy this condition along the
whole density interval $(0, \rho_{0})$, but only outside a certain 
temperature-dependent region $(\rho_{s1}(\beta),\rho_{s2}(\beta))$. 
For $\rho=\rho_{s1}(\beta)$ or $\rho=\rho_{s2}(\beta)$ the quantity $f$ 
vanishes, and consequently the compressibility diverges. As $\beta$ changes,
$\rho_{s1}(\beta)$ and $\rho_{s2}(\beta)$ give then respectively the low- 
and the high- density branch of the spinodal curve predicted by the theory. 
The fact that $f$ becomes negative for 
$\rho_{s1}(\beta)<\rho <\rho_{s2}(\beta)$ 
not only implies that the theory behaves unphysically
in this interval, but it also gives rise to an analytical instability 
which would make the numerical integration of the PDE~(\ref{pde}) impossible,
if one tried to determine the solution over the whole interval $(0,\rho_{0})$
even below the critical temperature. 
Therefore, the region bounded by the spinodal has been excluded from 
the integration of 
Eq.~(\ref{pde}). Specifically, as soon as it is found that $f$ changes sign, 
so that for a certain density $\widetilde{\rho}$ one has 
$f(\widetilde{\rho}, \beta)<0$, the integration is restricted to the interval
$(0,\widetilde{\rho}-\Delta\rho)$ or $(\widetilde{\rho}+\Delta\rho,\rho_{0})$
respectively for $\widetilde{\rho}<\rho_{c}$ or $\widetilde{\rho}>\rho_{c}$,
where $\Delta \rho$ is the spacing of the density grid. Within the precision
of the numerical discretization, one has 
$\rho_{s1}=\widetilde{\rho}-\Delta\rho$ 
(or $\rho_{s2}=\widetilde{\rho}+\Delta\rho$) and the further boundary 
conditions
\begin{equation}
u(\rho_{si},\beta)=u_{S}(\rho_{si}) 
\mbox{\hspace{1cm} $i=1,2,\mbox{\hspace{0.3cm}} \beta>\beta_{c}\, ,$}
\label{boundspin}
\end{equation}
where $\beta_{c}$ is the critical inverse temperature and $u_{S}(\rho)$ 
is the value of the internal energy per unit volume when the compressibility
at density $\rho$ diverges. This can be determined by setting $f=0$
in Eq.~(\ref{f2}) and solving for $\gamma_{1}$ as a function of $\rho$ and
$\gamma_{2}$. If Eqs.~(\ref{x}) and~(\ref{y}) are substituted into 
Eq.~(\ref{msa}), an equation for $\gamma_{2}$ is obtained that allows one
to determine the value of $\gamma_{2}$ when $1/\chi_{\rm red}=0$
for a certain $\rho$. Solving Eq.~(\ref{gamma22}) with respect to $u$ then 
yields $u_{S}(\rho)$. 

Once the internal energy per unit volume $u$ has been determined from 
Eq.~(\ref{pde}), the pressure $P$ and the chemical potential $\mu$ are 
obtained by integration with respect to $\beta$ via the relations 
$\partial (\beta P)/\partial \beta =-u+\rho \, \partial u/\partial \rho$,
$\partial (\beta \mu)/\partial \beta =\partial u/\partial \rho$. Thanks to
the self-consistency of the theory, this route to the thermodynamics 
is equivalent to integrating the inverse compressibility with respect 
to $\rho$, but it does not require one to circumvent the forbidden region 
in order to reach the high-density branch of the subcritical isotherms. 

\section{Results}

The numerical integration of the PDE~(\ref{pde}) with the initial 
condition~(\ref{init}) and the boundary 
conditions~(\ref{boundlow})--(\ref{boundspin}) has been performed on
a density grid with $\Delta \rho=10^{-3}$--$10^{-4}$. At the beginning of 
the integration the temperature step $\Delta \beta$ was usually set at
$\Delta \beta=2\times 10^{-5}$--$10^{-5}$. As the temperature approaches
its critical value, $\Delta \beta$ can be further decreased if one wishes
to get very close to the critical point, and then gradually expanded back. 
The integration was usually carried down to $\beta\simeq 2.4 \beta_{c}$.
The inverse range parameter of the attractive tail in Eq.~(\ref{yuk}) has 
been set at $z=1.8$. For this value of $z$ several 
simulations~\cite{henderson,smit,lomba} and 
theoretical~\cite{caccamoyuk,konior2,konior3} predictions have already been
reported in the literature. Fig.~1 shows the SCOZA results for the 
compressibility factor $Z=P/(\rho k_{B}T)$ along two different isotherms,
corresponding to $T=2$ and $T=1.5$, together with the MC simulation results
by Henderson and coworkers~\cite{henderson}. The agreement is very good both 
at low and high density. The compressibility factors are also reported in 
Tab.~1, together with those obtained by the LOGA/ORPA via the internal energy 
route~\cite{konior3}, which is the one that gives the best agreement with
the simulation results. 
It can be seen that for the non-critical states reported here the SCOZA and 
the energy route of the LOGA/ORPA are very close to each other. 
In Tab.~2 the predictions 
for the chemical potential and the reduced compressibility are compared to the
data from the MC simulations performed in this work. The internal energy per
particle is reported in Tab.~3, where again the LOGA/ORPA results 
are also shown.
The critical point predicted by the theory has been located by the vanishing
of the inverse compressibility $1/\chi_{\rm red}$. No extrapolation procedure 
to $1/\chi_{\rm red}=0$ is necessary, since the algorithm adopted here allows
one to get as close as desired to the critical singularity. As mentioned
in Sec.~2, below the critical temperature $T_{c}$ the theory yields a 
spinodal curve. The coexistence curve must be determined by a Maxwell 
construction, i.e. by imposing the equilibrium conditions 
$\mu (\rho_{g}, T)=\mu (\rho_{l},T)$, $P(\rho_{g},T)=P(\rho_{l},T)$ for the
densities $\rho_{g}$, $\rho_{l}$ of the gas and liquid phases at coexistence
at a temperature $T$. In comparing our results for the critical 
point and the coexistence curve with the available
simulation data, we found that the two simulations for the phase diagram 
of the system under study already reported in the 
literature~\cite{smit,lomba} do not agree very well with each other. We then
performed a new set of simulations using the MC-FSS method summarized in 
Appendix~B. The SCOZA and the simulation results for the critical point are
compared in Tab.~4, which also shows the predictions of other 
theories~\cite{caccamoyuk}. 
It can be seen that the agreement between the SCOZA and the present simulation
is remarkably good: the error in the critical density and temperature is
respectively slightly more and slightly less than 0.6\%. The SCOZA and the 
simulation coexistence curve in the temperature--density plane are compared
in Fig.~2. A similar comparison in the temperature--internal energy and in 
the temperature--chemical potential plane is shown respectively in Fig.~3 and 
in Fig~4. In every case the SCOZA agrees very well with the simulation.
It can be also appreciated that in the SCOZA the coexistence curve extends up
to the critical point, while, as already observed in the Introduction, 
this is not always the case with other theories. In Tab.~4 and in Figs.~2--4 
we have also reported the predictions of the simpler version of the SCOZA 
mentioned in Sec.~2, in which the direct correlation function outside the
repulsive core is given by just one Yukawa tail, and the thermodynamics 
of the hard-sphere gas is described in the PY approximation. It can be 
observed that even for not very short-ranged interaction the treatment of 
the repulsive contribution considerably affects the phase diagram predicted
by the theory, the two-Yukawa SCOZA sensibly improving over the one-Yukawa
version. The behavior of the SCOZA in the critical region has been 
studied both analytically~\cite{scozacrit} and numerically~\cite{scoza1}. This 
investigation has shown that above the critical temperature the SCOZA yields 
the same critical exponents as the mean spherical approximation (MSA), i.e.
$\gamma=2$, $\delta=5$, $\alpha=-1$,
where the usual notation for the critical exponents has been used. On the
other hand, on the coexistence curve the critical exponents are neither 
spherical nor classical, and one finds $\gamma^{\prime}=7/5$, 
$\alpha^{\prime}=-1/10$, $\beta=7/20$ (here of course $\beta$ is the critical
exponent that gives the curvature of the coexistence curve near the critical
point). Although these results were determined in the case of a 
nearest-neighbor attractive lattice gas, we expect them to hold also in the
continuum case. Fig.~5 shows the reduced compressibility of the HCYF for
$T>T_{c}$ and $\rho=\rho_{c}$ as a function of the reduced 
temperature $t=(T-T_{c})/T_{c}$ on a log-log plot. Also shown is the
correspondent {\it effective exponent} $\gamma_{\rm eff}$, defined as the
local slope of the plot. It can be seen that $\gamma_{\rm eff}$  
eventually saturates at $\gamma=2$, thus signaling the onset of a MSA-like
power-law behavior, but the asymptotic regime can be detected only at very 
small reduced temperature ($t\sim 10^{-6}$). This is the same scenario 
previously found in the nearest-neighbor lattice gas. For the HCYF, the 
crossover is controlled as expected by the inverse range parameter $z$.
It has been verified that as the interaction becomes longer and longer ranged, 
the asymptotic regime is further pushed to smaller and smaller values of 
the reduced temperature $t$~\cite{borge}. 

\section{Conclusions}

We have studied the thermodynamics and the phase diagram of the HCYF using 
both the SCOZA and MC simulations supplemented by a finite-size scaling 
analysis.
The comparison between theory and simulation results shows that the SCOZA 
yields
both very good overall thermodynamics and a remarkably accurate coexistence 
curve up to the critical point. The version of the 
SCOZA considered here takes 
into account the hard-sphere contribution to the direct correlation function
outside the repulsive core, and sensibly improves over the simpler 
one-Yukawa version, in which the hard-sphere gas is described as in 
the PY approximation. On the other hand, as stated in Sec.~2,
here (as well as in the simpler version just mentioned) consistency has been 
enforced between the internal energy and the compressibility route, but not
between the virial route and either of the above. We think that the further 
development of making the theory fully self-consistent by taking also the
virial route into account is worth pursuing, since we anticipate 
that the present version of the SCOZA will yield liquid-state 
pressures
from the virial theorem that are not significantly better than those
obtained using the virial theorem with the LOGA/ORPA $g(r)$. 
We defer a detailed examination of this issue to a later study.
In this respect it is
worth mentioning an investigation of the HCYF along the lines considered 
here~\cite{tau}, where some results for the critical parameters 
were reported taking into account all the three routes to the thermodynamics
although, as explicitly pointed out by the authors, the SCOZA equations 
were studied in an approximate fashion, and no attempt to determine the phase 
diagram was made. 

Although dealing with a Yukawa potential entails certain analytical
simplifications in implementing the SCOZA, such an approach can be applied 
to any kind of tail potential. It should also be pointed out that 
the idea of using the requirement of self-consistency to get 
a closed theory of thermodynamics and correlations is pertinent not only 
to the realm of simple fluids or lattice gases, but has also proven 
to be a powerful tool in the study of a system of spins with continuous
symmetry~\cite{hoye3}, and of a site-diluted~\cite{kierlik1} 
or random-field~\cite{kierlik2} Ising model. 
\vspace{1cm}

D.P. gratefully acknowledges the support of the Division of Chemical Sciences,
Office of Basic Energy Sciences, Office of Energy Research, U.S.~Department
of Energy. G.S. gratefully acknowledges the support of the National Science
Foundation. N.B.W. acknowledges support from the EPSRC 
(grant number GR/L91412), and the Royal Society of Edinburgh. 

\appendix
\section{Waisman parameterization of $c_{\rm HS}(r)$}

In this Appendix we recall the procedure that allows one to determine 
analytically the amplitude $K_{1}$ and the inverse range $z_{1}$ of the
direct correlation function $c_{\rm HS}(r)$ of the hard-sphere gas 
in the Waisman parameterization~(\ref{waisman}). The relevant equations 
are reported in Ref.~\cite{oneyuk}, which will be referred here as IV. 
Both $z_{1}$ and $K_{1}$ are conveniently expressed in terms of two quantities
$V_{0}$, $V_{1}$ which are formally analogous to $U_{0}$, $U_{1}$ and 
$W_{0}$, $W_{1}$ introduced in Eqs.~(\ref{gamma1}),~(\ref{gamma2}).
>From Eq.~(IV.2.32 a) it is found that $z_{1}$ is given by
\begin{equation}
z_{1}=\frac{2}{q-f^{2}}
\left[\, (V_{0}+f^{2}-q)f+\sqrt{(V_{0}+f^{2}-q)V_{0}\, q}\, \right] \, ,
\label{z1}
\end{equation}
where $f$ and $q$ are defined in Eqs.~(\ref{f}),~(\ref{q}). 
The expression of the amplitude $K_{1}$ is the same as in Eqs.~(\ref{k1})
with $U_{0}$, $U_{1}$ replaced by $V_{0}$, $V_{1}$:
\begin{equation}
K_{1} = \frac{2(z_{1}+2)^{2}\sigma_{1}^{2}}{3\xi z_{1}^{2}}\, V_{0}
\left[\, \frac{V_1}{V_0}-\alpha_{1}\right]^{2} \, ,
\label{k1a}
\end{equation}
where $\xi$ is the packing fraction $\xi=\pi \rho/6$ and 
$\alpha_{1}$ is a function of $z_{1}$ given by Eq.~(\ref{alpha}). 
The ratio $V_{1}/V_{0}$ can be expressed as a function of $V_{0}$ 
by Eqs.~(IV.2.24) and~(IV.2.26). One has
\begin{equation}
\frac{V_{1}}{V_{0}}=2-\sqrt{q}-\frac{1}{2V_{0}\sqrt{q}}
\left[\, (V_{0}+f^{2}-q)(V_{0}+f^{2})+\frac{1}{4}z_{1}^{2}(q-f^{2})\, 
\right] \, .
\label{vrat}
\end{equation}
To obtain the explicit expressions of $z_{1}$ and $K_{1}$ as functions of the
density, one must then feed into Eqs.~(\ref{z1})--(\ref{vrat}) the expression
of $V_{0}$. This depends on the contact value of the radial distribution
function $y_{0}\equiv g(r\! =\! 1^{+})$ via Eq.~(IV.2.32 b):
\begin{equation}
V_{0}=6\xi \, y_{0}-f^{2}+1 \, .
\label{v}
\end{equation}
For a hard-sphere gas $y_{0}$ can be determined from the equation of state
via the virial equation:
\begin{equation}
\frac{\beta P}{\rho}=1+4\xi \, y_{0} \, .
\label{vir}
\end{equation}
The requirement that both the virial and the compressibility route to the
thermodynamics must give the Carnahan-Starling equation of state is then
satisfied if the Carnahan-Starling pressure and compressibility are
substituted respectively in Eq.~(\ref{vir}) and Eq.~(\ref{f}). 
Eqs.~(\ref{v}) and~(\ref{vrat}) then yield $V_{0}$ and $V_{1}/V_{0}$
as a function of density. From Eqs.~(\ref{z1}) and~(\ref{k1a}) we finally
get $z_{1}$ and $K_{1}$. 

\section{Simulation details}

The principal aspects of the simulation and finite-size scaling
techniques employed in this work have previously been detailed
elsewhere in the context of a similar study of the Lennard-Jones
fluid. Accordingly we confine our description to the barest essentials
and refer the reader to reference \cite{wilding} for a fuller
account of our methods.

The Monte-Carlo simulations were performed using a Metropolis algorithm
within the grand canonical ensemble \cite{adams}. The MC scheme
comprises only particle transfer (insertion and deletion) steps,
leaving particle moves to be performed implicitly as a result of
repeated transfers. To simplify identification of particle interactions
a linked-list scheme was employed. This involves partitioning  the
periodic simulation space of volume $L^3$ into $m^3$ cubic cells, each
of side the cutoff $r_c$. This strategy ensures that interactions
emanating from particles in a given cell extend at most to particles in
the $26$ neighbouring cells. 

In our Yukawa system the potential was cutoff at a radius
$r_c=3.0\sigma$, and a correction term was applied to the internal
energy to compensate for the trunction. System sizes having $m=3,4,5,6$
and $7$ were studied, corresponding (at coexistence) to average
particle numbers of approximately $230,540,1050,1750$ and $2900$
respectively. For the $m=3,4$ and $5$ system sizes, equilibration
periods of $10^5$ Monte Carlo transfer attempts {\em per cell} (MCS)
were utilised, while for the $m=6$ and $m=7$ system sizes up to
$2\times10^6$ MCS were employed. Sampling frequencies ranged from $20$
MCS for the $m=3$ system to $150$ MCS for the $m=7$ system. The total
length of the production runs was also dependent upon the system size.
For the $m=3$ system size, $1\times10^7$ MCS were employed, while for
the $m=7$ system, runs of up to $6\times 10^7$ MCS were necessary. 

In the course of the simulations, the observables recorded were the
particle number density $\rho=N/V$ and the energy density $u=E/V$. The
joint distribution $p_L(\rho ,u)$ was accumulated in the form of a
histogram. In accordance with convention, we express $\rho$ and $u$ in
reduced units: $\rho^\ast = \rho \sigma ^3,  u^\ast = u \sigma ^3$. To
allow us to explore efficiently the phase space of the model, we
employed the histogram reweighting technique \cite{ferrenberg}. This
method allows histogram accumulated at one set of model parameters to
be reweighted to provide estimates appropriate to another set of
not-too-distant model parameters.

To facilitate study of the subcritical coexistence region, the
multicanonical preweighting technique \cite{berg} was employed. This
technique allows one to circumvent the problems of metastability and
nonergodicity that would otherwise arise from the large free energy
barrier separating the coexisting phases. Details of this technique and
its implementation in the fluid context are described in reference
\cite{wilding}. 

The critical point parameters were estimated using finite-size scaling
technique as described in \cite{wilding}. This involves matching the
distribution function of the ordering operator to the independently
known universal critical point form appropriate for the Ising
universality class. The ordering operator is defined as ${\cal
M}\propto(\rho^\ast+su^\ast)$, where $s$  is a non-universal ``field
mixing'' parameter, which is finite in the absence of particle-hole
symmetry, and which is chosen to ensure that $p({\cal M})$ is symmetric
in ${\cal M}$. The estimate of the apparent critical
temperature obtained by this matching procedure is, however, subject to
errors associated with corrections to finite-size scaling. To deal with
this, we extrapolate to the thermdodynamic limit using the known
scaling properties of the corrections, which are expected to diminish
(for sufficiently large system sizes) like $L^{-\theta/\nu}$
\cite{wilding}, where $\theta$ is the correction to scaling exponent
and $\nu$ is the correlation length exponent. The extrapolation has
been performed using a least squares fit to the data for the four
largest system sizes. The results of the extrapolation are shown in
figure A1, from which we estimate $T_c=1.212(2)$. The associated
estimate for the critical density is $\rho^\ast_c=0.312(2)$.

\newpage
\begin{center}
TABLE 1

\vspace*{1truecm}
\begin{tabular}{ccccc}  \hline \hline
\makebox[1.5cm]{$T^{\ast}$} & 
\makebox[1.5cm]{$\rho^{\ast}$} & 
\makebox[1.5cm]{${\rm MC}^{\dagger}$} & 
\makebox[1.5cm]{SCOZA} &
\makebox[3cm]{${\rm LOGA/ORPA}_{{\rm en}}^\diamond$} \\
\hline
$\infty $  & 0.4 & 2.52 & 2.518 & 2.518 \\
$\infty $  & 0.6 & 4.22 & 4.283 & 4.283 \\
$\infty $  & 0.8 & 7.65 & 7.750 & 7.750 \\
   2.0     & 0.4 & 1.08 & 1.120 & 1.118 \\
   2.0     & 0.6 & 2.04 & 1.977 & 1.974 \\   
   2.0     & 0.8 & 4.27 & 4.433 & 4.432 \\    
   1.5     & 0.4 & 0.69 & 0.667 & 0.663 \\
   1.5     & 0.6 & 1.21 & 1.220 & 1.214 \\
   1.5     & 0.8 & 3.31 & 3.333 & 3.330 \\
\hline \hline
\end{tabular}
\end{center}
\vspace*{1truecm}
Compressibility factor $PV/Nk_{B}T$ for the hard-sphere + Yukawa fluid
($z=1.8$). Density and temperature are in reduced units 
$\rho^{\ast}\!=\!\rho\sigma^{3}$, $T^{\ast}\!=\!k_{B}T/\epsilon$, 
where $\sigma$ 
is the hard-sphere diameter and $\epsilon$ is the strength of the attractive 
potential. 
$\dagger$: Monte Carlo data from Ref.~\cite{henderson}.
$\diamond$: LOGA/ORPA-energy route results from Ref.~\cite{konior3}.

\newpage
\begin{center}
TABLE 2
\vspace*{1truecm}

\begin{tabular}{ccrr|ll}  
\hline 
  &   & \multicolumn{2}{c}{\hspace{0.3cm}$\mu/k_{B}T$} 
& \multicolumn{2}{c}{$\chi_{\rm red}$}  \\ 
\hline \hline  
\makebox[1.5cm]{$T^{\ast}$} &
\makebox[1.5cm]{$\rho^{\ast}$} &
\makebox[1.5cm]{\hspace{0.1cm} ${\rm MC}^{\dagger}$} &
\makebox[1.6cm]{\hspace{0.2cm} SCOZA} & 
\makebox[1.8cm]{${\rm MC}^{\dagger}$} &
\makebox[2cm]{\hspace{-0.8cm} SCOZA}  \\
\hline
 $\infty $  & 0.4 &   1.736(2)  &   1.7316 & 0.1958(2)  & 0.19744 \\
 $\infty $  & 0.6 &   4.833(2)  &   4.8147 & 0.0848(5)  & 0.08721 \\
    2.0     & 0.4 & --0.936(2)  & --0.9396 & 0.4992(8)  & 0.50439 \\
    2.0     & 0.6 &   0.515(2)  &   0.5003 & 0.1594(5)  & 0.15976 \\
    1.5     & 0.4 & --1.823(2)  & --1.8258 & 0.968(3)   & 1.0150  \\
    1.5     & 0.6 & --0.905(2)  & --0.9294 & 0.2217(5)  & 0.22147 \\
\hline \hline
\end{tabular}
\end{center}
\vspace*{1truecm}
Chemical potential $\mu$ and reduced compressibility $\chi_{\rm red}$ 
of the hard-sphere Yukawa 
fluid $(z=1.8)$. Density and temperature are in reduced units.
$\dagger$: Monte Carlo simulation performed in this work. 
The numbers in brackets give the error in the last figure.

\newpage
\begin{center}
TABLE 3

\vspace*{1truecm}

\begin{tabular}{cccccc}  
\hline \hline  
\makebox[1.5cm]{$T^{\ast}$} &
\makebox[1.5cm]{$\rho^{\ast}$} &
\makebox[1.5cm]{${\rm MC}^{\dagger}$} &
\makebox[1.5cm]{${\rm MC}^{\ddagger}$} & 
\makebox[1.5cm]{SCOZA} &
\makebox[3cm]{${\rm LOGA/ORPA}^{\diamond}$} \\
\hline
 $\infty $  & 0.4 & --2.495 & --2.516(2) & --2.517 & --2.517 \\
 $\infty $  & 0.6 & --3.975 & --4.002(2) & --4.002 & --4.002 \\
 $\infty $  & 0.8 & --5.573 &            & --5.611 & --5.611 \\
    2.0     & 0.4 & --2.583 & --2.595(2) & --2.583 & --2.574 \\
    2.0     & 0.6 & --4.030 & --4.036(2) & --4.030 & --4.026 \\
    2.0     & 0.8 & --5.622 &            & --5.620 & --5.618 \\
    1.5     & 0.4 & --2.622 & --2.640(2) & --2.623 & --2.602 \\
    1.5     & 0.6 & --4.051 & --4.053(2) & --4.043 & --4.036 \\
    1.5     & 0.8 & --5.630 &            & --5.623 & --5.621 \\
    1.0     & 0.6 & --4.073 &            & --4.097 & --4.065 \\
    1.0     & 0.8 & --5.635 &            & --5.631 & --5.628 \\
\hline \hline
\end{tabular}
\end{center}
\vspace*{1truecm}
Internal energy per particle of the hard-sphere Yukawa fluid. 
All quantities are in reduced units.
$\dagger$: Monte Carlo simulation of Ref.~\cite{henderson}.
$\ddagger$: Monte Carlo simulation performed in this work. The number in 
brackets give the error in the last figure.
$\diamond$: LOGA/ORPA. The entries for $T^{\ast}=\infty$ are from this work,
the rest are from Ref.~\cite{konior2}. 

\newpage

\begin{center}
TABLE 4

\vspace*{1truecm} 
\noindent
\begin{tabular}{cccccccc}  \hline \hline
\makebox[1.2cm]&
\makebox[1.5cm]{\hspace{0.1cm}${\rm MC}^{\circ}$} &
\makebox[1.5cm]{\hspace{0.1cm}${\rm MC}^{\diamond}$} &
\makebox[1.5cm]{\hspace{0.1cm}${\rm MC}^{\dagger}$} & 
\makebox[2cm]{\hspace{0.5cm} ${\rm SCOZA}_{\rm 1-Yuk}^{\!\!\ddagger}$} & 
\makebox[2cm]{SCOZA} &
\makebox[1.5cm]{${\rm HMSA}^\bullet $} & 
\makebox[1.5cm]{${\rm MHNC}^\bullet $} \\
\hline

\makebox[1.2cm]{$\rho_c^{\ast}$} & 0.294 & 0.313 & 0.312(2) & 0.308 
& 0.314 & 0.36 & 0.28    \\  
\makebox[1.2cm]{$T_c^{\ast}$} & 1.192 & 1.178 & 1.212(2) & 1.201 
& 1.219 & 1.25 & 1.21  \\ \hline \hline
  
\end{tabular}
\end{center}
\vspace*{1truecm} 
\noindent
Critical density and temperature (in reduced units) for the hard-sphere 
Yukawa fluid. 
$\circ$:~MC simulation of Ref.~\cite{smit}.
$\diamond$:~MC simulation of Ref.~\cite{lomba}
$\dagger$:~MC simulation performed in this work.
$\ddagger$:~SCOZA with 1-Yukawa $c(r)$ (see text). 
$\bullet$:~from Ref.~\cite{caccamoyuk}

\newpage

\begin{center}
FIGURE CAPTIONS
\end{center}

\vspace*{1.5cm}
\noindent
\begin{description}
\item[Fig. 1] Compressibility factor $Z= P/(\rho k_{B}T)$ of the hard-sphere 
Yukawa fluid ($z=1.8$) as a function of the reduced density $\rho^{\ast}$ 
along two isotherms at reduced temperature $T^{\ast}=2$ (upper curve) 
and $T^{\ast}=1.5$ (lower curve). 
Full curve: SCOZA. Squares: MC simulation results~\cite{henderson}.
\item[Fig. 2] Coexistence curve of the hard-sphere Yukawa fluid ($z=1.8$)
in the density--temperature plane. Density and temperature are in reduced 
units. Full curve: SCOZA. Dashed curve: SCOZA with a one-Yukawa 
direct correlation function $c(r)$ (see text). Squares: MC results 
(this work). 
\item[Fig. 3] Coexistence curve of the hard-sphere Yukawa fluid in the 
internal energy--temperature plane. $E^{\ast}/N$ is the internal 
energy per particle in reduced units. Notation as in Fig.~2.
\item[Fig. 4] Coexistence curve of the hard-sphere Yukawa fluid in the
temperature--chemical potential plane. All quantities are in reduced 
units. Notation as in Fig.~2. 
\item[Fig. 5] Log-log plot of the reduced compressibility $\chi_{\rm red}$
of the hard-sphere Yukawa fluid ($z=1.8$) on the critical isochore as 
a function of the reduced temperature $t=(T-T_{c})/T_{c}$ according to the 
SCOZA (a) and effective exponent $\gamma_{\rm eff}$, defined as 
$\gamma_{\rm eff}=-d(\log \chi_{\rm red})/d(\log t)$ (b). 
\item[Fig. A1] The apparent critical temperature, (as defined
by the matching condition described in the text), plotted as a function of
$L^{-(\theta+1)/\nu}$, with $\theta=0.54$ and $\nu=0.629$. The
extrapolation of the least squares fit to infinite volume yields the
estimate $T_c^\ast=1.212(2)$.

\end{description}
\newpage


\begin{thebibliography}{99}
%
\bibitem{barker} Barker, J. A., and Henderson, D., 1967, {\it J. Chem. Phys.}
{\bf 47}, 2856, 4714. These are parts~I and~II of a study entitled
{\it Perturbation Theory and Equation of State for Fluids}, with subtitles
{\it The Square-Well Potential} and {\it A Successful Theory of Fluids} 
respectively.
%
\bibitem{mansoori} See, e.g., Mansoori, G. A., and Canfield, F. B., 1969, 
{\it J. Chem. Phys.}
{\bf 51}, 4958; Rasaiah, J., and Stell, G., 1970, {\it Mol. Phys.} {\bf 18},
249; Weeks, J. D., Chandler, D., and Andersen, H. C., 1971, 
{\it J. Chem. Phys.} {\bf 54}, 5237; {\bf 55}, 5421. 
%
\bibitem{hauge} Hauge, E. H., and Hemmer, P. C., 1966, {\it J. Chem. Phys.}
{\bf 45}, 223. 
%
\bibitem{caccamorev} Caccamo, C., 1996, {\it Phys. Reports} {\bf 274}, 1,
provides an excellent description of modern integral-equation 
theories of $g(r)$. 
%
\bibitem{rosenfeld} Rosenfeld, Y., and Ashcroft, N. W., 1979, 
{\it Phys. Rev A} {\bf 20}, 1208.
%
\bibitem{caccamoyuk} Caccamo, C., Giunta, G., and Malescio, G., 1995, 
{\it Mol. Phys.} {\bf 84}, 125.
%
\bibitem{caccamoljrhnc} Caccamo, C., 1995, {\it Phys. Rev. B} {\bf 51}, 3387.
%
\bibitem{caccamoljhmsa} Caccamo, C., Giaquinta, P. V., and Giunta, G., 1993,
{\it J. Phys.: Condens. Matter} {\bf 5}, B75
%
\bibitem{zerah} Zerah, G., and Hansen, J. P., 1985, {\it J. Chem. Phys.} 
{\bf 84}, 2336. 
%
\bibitem{scozacrit} H\o ye, J. S., Pini, D., and Stell, G., unpublished.
%
\bibitem{dickman} Dickman, R., and Stell, G., 1996, {\it Phys. Rev. Lett.}
{\bf 77}, 996.
%
\bibitem{scoza1} Pini, D., Stell, G., and Dickman, R., 1998, 
{\it Phys. Rev. E}, in press.
%
\bibitem{hoye1} H\o ye, J. S., and Stell, G., 1977, {\it J. Chem. Phys.}
{\bf 67}, 439. 
%
\bibitem{twoyuk3} H\o ye, J. S., and Stell, G., 1984, {\it Mol. Phys.}
{\bf 52}, 1071.  
%
\bibitem{scoza2} Pini, D., Stell, G., and H\o ye, J. S., submitted to
{\it Int. J. Thermophys.} 
%
\bibitem{wilding} Wilding, N. B., 1995, {\it Phys. Rev. E} {\bf 52},
602; Wilding, N.B. Bruce, A.D., 1992, {\em J. Phys. Condens. Matter}
{\bf 4}, 3087.
%
\bibitem{stell} Stell, G., 1971, {\it J. Chem. Phys.} {\bf 55}, 1485.
%
\bibitem{andersen} Andersen, H. C., and Chandler, D., 1972, 
{\it J. Chem. Phys} {\bf 57}, 1918.  
%
\bibitem{verlet} Verlet, L., and Weis, J. J., 1972, {\it Phys. Rev. A}
{\bf 5}, 939; Henderson, D., and Grundke, E. W., 1975, {\it J. Chem. Phys.}
{\bf 63}, 601.
%
\bibitem{waisman} Waisman, E., 1973, {\it Mol. Phys.} {\bf 25}, 45.
%
\bibitem{oneyuk} H\o ye, J. S., and Stell, G., 1976, {\it Mol. Phys.}
{\bf 32}, 195.  
%
\bibitem{blum} Henderson, D., Stell, G., and Waisman, E., 1975, 
{\it J. Chem. Phys.} {\bf 62}, 4247;
Henderson, D., and Blum, L., 1976, {\it Mol. Phys.} {\bf 32}, 1627. 
%
\bibitem{twoyuk1} H\o ye, J. S., Stell, G., and Waisman, E., 1976, 
{\it Mol. Phys.} {\bf 32}, 209.
%
\bibitem{twoyuk2} H\o ye, J. S., and Stell, G., 1984, {\it Mol. Phys.} 
{\bf 52}, 1057. 
%
\bibitem{konior1} Konior, J., and Jedrzejek, C., 1983, {\it Mol. Phys.} 
{\bf 48}, 219.
%
\bibitem{konior2} Konior, J., and Jedrzejek, C., 1985, {\it Mol. Phys.}
{\bf 55}, 187. 
%
\bibitem{konior3} Konior, J., 1989, {\it Mol. Phys.} {\bf 68}, 129.
%
\bibitem{ames} Ames, W. F., 1977, {\it Numerical Methods for Partial 
Differential Equations} (New York: Academic Press).
%
\bibitem{packing} Stell, G., and Penrose, O., 1983, {\it Phys. Rev. Lett.}
{\bf 51}, 1397.  
%
\bibitem{henderson} Henderson, D., Waisman, E., Lebowitz, J. L., Blum, L.,
1978, {\it Mol. Phys.} {\bf 35}, 241.  
%
\bibitem{smit} Smit, B., and Frenkel, D., 1991, {\it Mol. Phys.} 
{\bf 74}, 35.
%
\bibitem{lomba} Lomba, E., and Almarza, N. G., 1994, {\it J. Chem. Phys.}
{\bf 100}, 8367.
%
\bibitem{borge} Borge, A., and H\o ye, J. S., 1998, {\it J. Chem. Phys.},
in press.
%
\bibitem{tau} Tau, M., and Reatto, L., 1985, {\it J. Chem. Phys.} 
{\bf 83}, 1921. 
%
\bibitem{hoye3} H\o ye, J. S., and Stell, G., 1997, {\it Physica A}, 
{\bf 244}, 176; {\bf 247}, 497.
%
\bibitem{kierlik1} Kierlik, E., Rosinberg, M. L., and Tarjus, G., 1997,
{\it J. Stat. Phys.} {\bf 89}, 215. 
%
\bibitem{kierlik2} Kierlik, E., Rosinberg, M. L., and Tarjus, G., preprint
%
\bibitem{adams} Adams, D. J., 1975, {\it Mol. Phys.} {\bf 29}, 307.
%
\bibitem{ferrenberg} Ferrenberg, A. M., and Swendsen, R. H., 1988,
{\it Phys. Rev. Lett.} {\bf 61}, 2635; {\it ibid} 1989, {\bf 63}, 1195.
%
\bibitem{berg} Berg, B., and Neuhaus, T., 1992, 
{\it Phys. Rev. Lett.} {\bf 68}, 9.
%
%
%
%
%
%
%
%
%
%
\end{thebibliography}
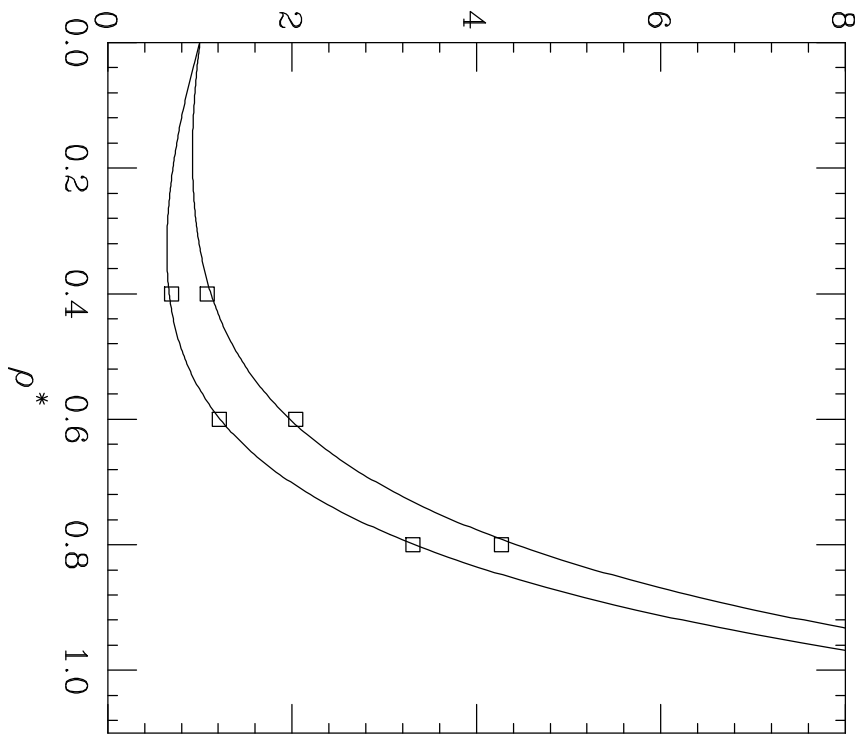
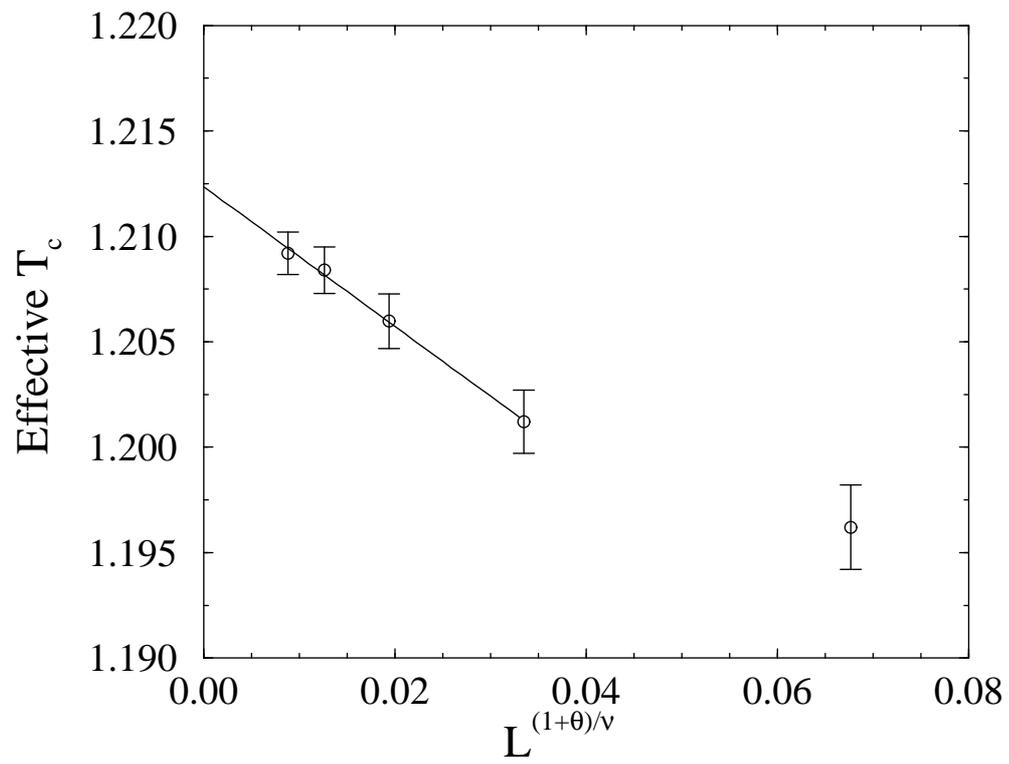
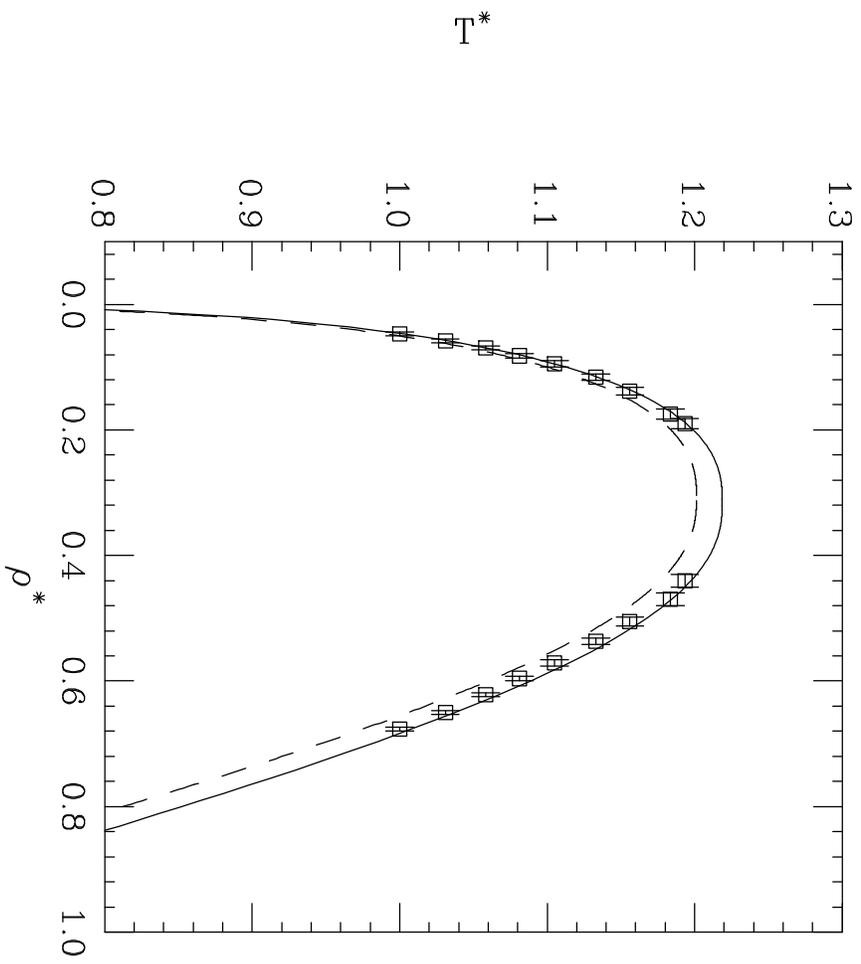
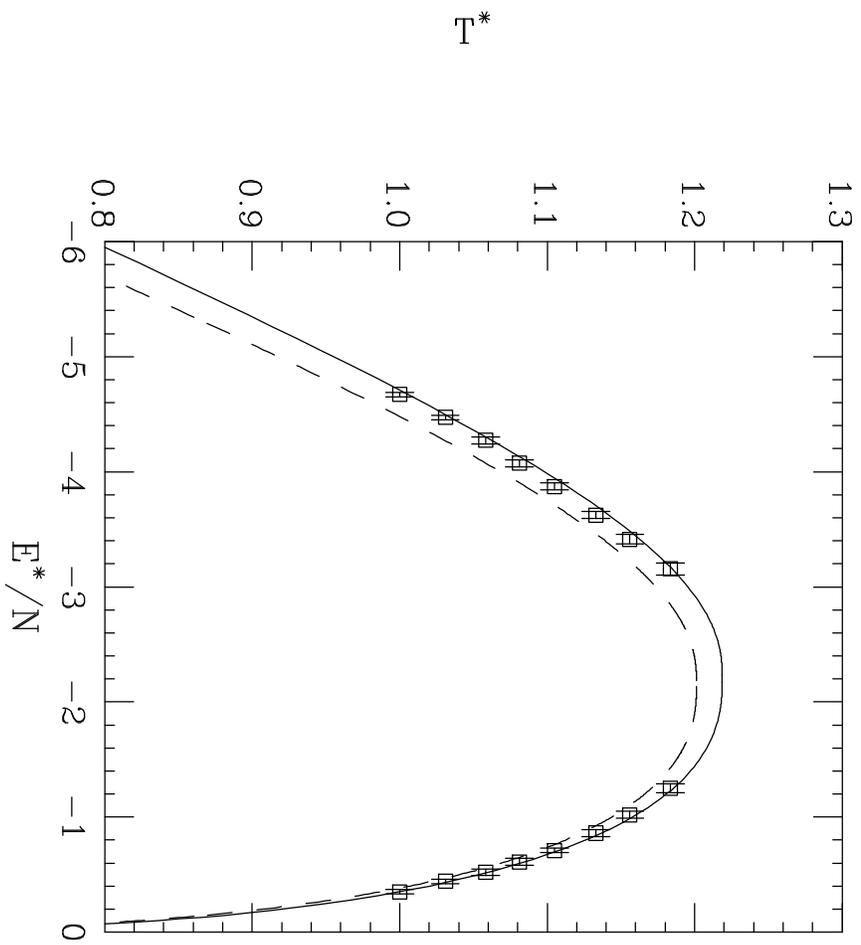
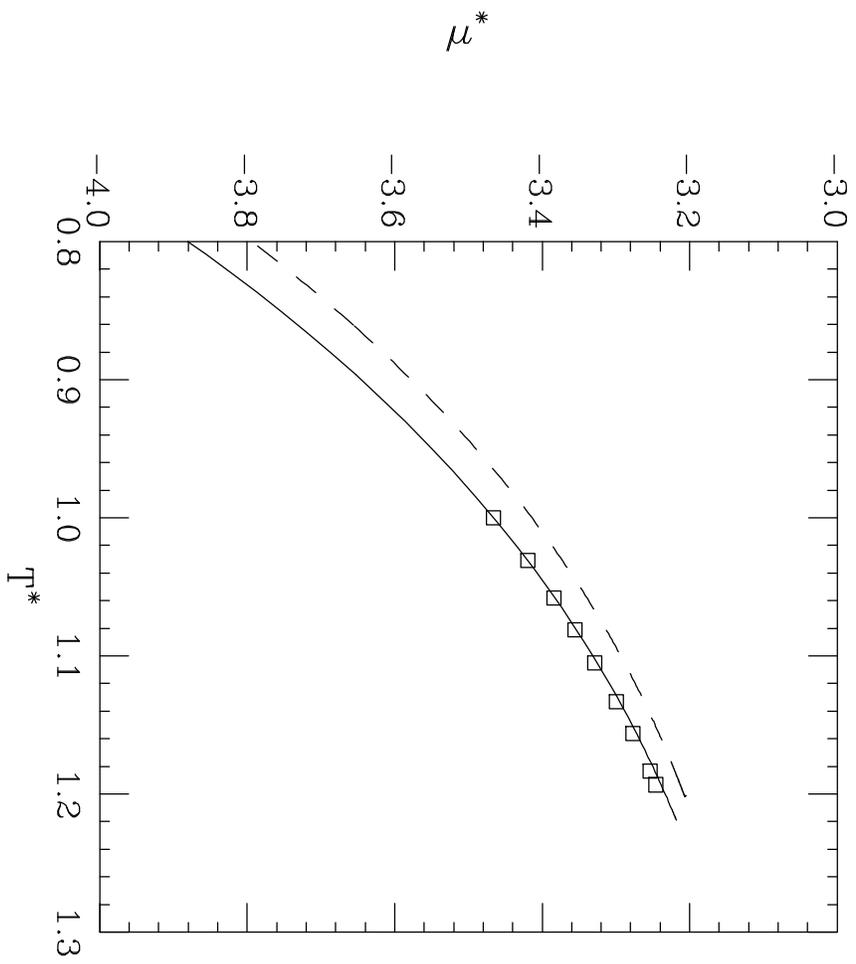
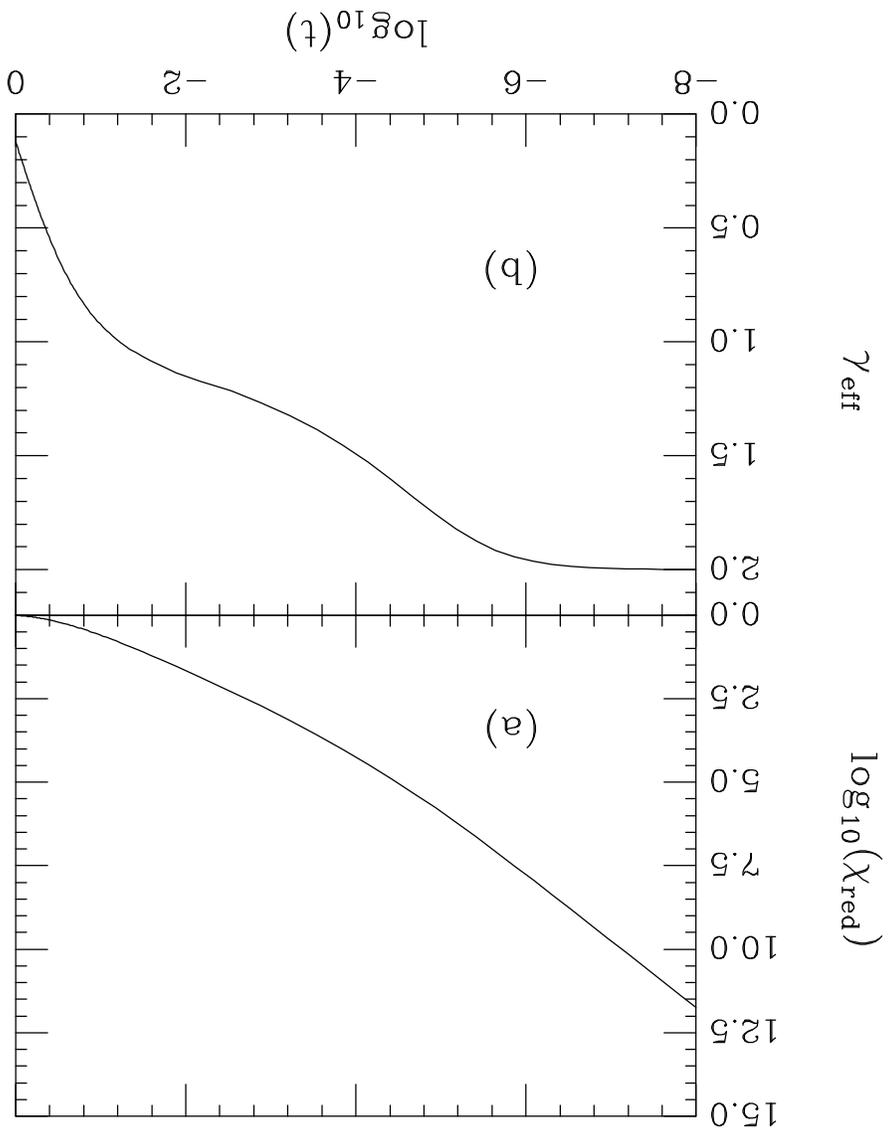
\end{document}